\documentclass[manuscript, screen, natbib=true]{acmart}
\usepackage{courier}
\usepackage{array}
\usepackage{float}
\usepackage{setspace}
\usepackage{xcolor}
\definecolor{green}{RGB}{0,150,0}
\usepackage{graphicx}
\usepackage{epstopdf}
\usepackage{subfigure}
\usepackage{natbib}

\setcopyright{acmlicensed}
\copyrightyear{2024}
\acmYear{2024}
\acmDOI{XXXXXXX.XXXXXXX}

\acmConference[ACM-BCB]{The 15th ACM Conference on Bioinformatics, Computational Biology, and Health Informatics}{Nov. 22--25, 2024}{Shenzhen, Guangdong Province, PR China}
\acmISBN{978-1-4503-XXXX-X/18/06}

\begin{document}

\title[Peptide Sequencing Via Protein Language Models]{Peptide Sequencing Via Protein Language Models}


\author{Thuong Le Hoai Pham}
\affiliation{%
 \institution{University of Texas at Arlington}
 \city{Arlington}
 \state{Texas}
 \country{USA}
}

\author{Jillur Rahman Saurav}
\authornote{These authors contributed equally to this research.}
\author{Aisosa A. Omere}
\authornotemark[1]
\author{Calvin J. Heyl}
\authornotemark[1]
\affiliation{%
 \institution{University of Texas at Arlington}
 \city{Arlington}
 \state{Texas}
 \country{USA}
}

\author{Mohammad Sadegh Nasr}
\affiliation{%
 \institution{University of Texas at Arlington}
 \city{Arlington}
 \state{Texas}
 \country{USA}
}

\author{Cody Tyler Reynolds}
\affiliation{%
 \institution{University of Texas at Arlington}
 \city{Arlington}
 \state{Texas}
 \country{USA}
}

\author{Jai Prakash Yadav Veerla}
\affiliation{%
 \institution{University of Texas at Arlington}
 \city{Arlington}
 \state{Texas}
 \country{USA}
}

\author{Helen H Shang}
\affiliation{%
 \institution{UCLA Health}
 \city{Los Angeles}
 \state{California}
 \country{USA}
}

\author{Justyn Jaworski}
\affiliation{%
 \institution{University of Texas at Arlington}
 \city{Arlington}
 \state{Texas}
 \country{USA}
}

\author{Alison Ravenscraft}
\affiliation{%
 \institution{University of Texas at Arlington}
 \city{Arlington}
 \state{Texas}
 \country{USA}
}

\author{Joseph Anthony Buonomo}
\authornote{To whom correspondence should be addressed (jacob.luber@uta.edu \& joseph.buonomo@uta.edu).}
\email{joseph.buonomo@uta.edu}
\affiliation{%
 \institution{University of Texas at Arlington}
 \city{Arlington}
 \state{Texas}
 \country{USA}
}

\author{Jacob M. Luber}
\authornotemark[2]
\email{jacob.luber@uta.edu}
\affiliation{%
 \institution{University of Texas at Arlington}
 \city{Arlington}
 \state{Texas}
 \country{USA}
}

\renewcommand{\shortauthors}{Pham et al.}

\begin{abstract}
 We introduce a protein language model for determining the complete sequence of a peptide based on measurement of a limited set of amino acids. To date, protein sequencing relies on mass spectrometry, with some novel edman degregation based platforms able to sequence non-native peptides. Current protein sequencing techniques face limitations in accurately identifying all amino acids, hindering comprehensive proteome analysis. Our method simulates partial sequencing data by selectively masking amino acids that are experimentally difficult to identify in protein sequences from the UniRef database. This targeted masking mimics real-world sequencing limitations. We then modify and finetune a ProtBert derived transformer-based model, for a new downstream task predicting these masked residues, providing an approximation of the complete sequence. Evaluating on three bacterial \textit{Escherichia} species, we achieve per-amino-acid accuracy up to 90.5\% when only four amino acids ([KCYM]) are known. Structural assessment using AlphaFold and TM-score validates the biological relevance of our predictions. The model also demonstrates potential for evolutionary analysis through cross-species performance. This integration of simulated experimental constraints with computational predictions offers a promising avenue for enhancing protein sequence analysis, potentially accelerating advancements in proteomics and structural biology by providing a probabilistic reconstruction of the complete protein sequence from limited experimental data.
\end{abstract}

\ccsdesc[500]{Applied computing~Sequencing and genotyping technologies}
\ccsdesc[300]{Applied computing~Bioinformatics}
\ccsdesc[100]{Applied computing~Computational proteomics}

\keywords{Computational Biology, Protein Sequencing, High Performance Computing, Machine Learning, Language Modeling, Deep Learning }

\received{15 July 2024}

\maketitle

\section{Introduction}

Protein sequences are fundamental to understanding biological processes, disease mechanisms, and therapeutic developments \cite{lieu2020amino,maddocks2017modulating}. Despite significant advancements in genomics and proteomics, aided by machine learning (ML) techniques \cite{libbrecht2015machine, wen2020deep}, accurate and comprehensive protein sequencing remains a challenge in the field \cite{alfaro2021emerging}.

Protein sequencing methods primarily rely on techniques such as Edman degradation \cite{niall197336} and mass spectrometry (MS) \cite{hunt1986protein}, including liquid chromatography tandem mass spectrometry (LC-MS/MS) \cite{vogeser2007liquid}. While these methods have advanced our understanding of proteins, they face significant limitations in accurately identifying all amino acids in a sequence, particularly for complex or low-abundance proteins \cite{alfaro2021emerging}. These limitations often result in partially known sequences, hindering comprehensive proteome analysis.

Despite these advancements, protein sequencing still faces significant challenges, including high error rates, complex data interpretation, and technological limitations \cite{smith2024estimating, filius2024full, searle2024nanopore}. Overcoming these hurdles requires further advancements in sequencing technologies, sophisticated data processing algorithms, and improved experimental protocols to enhance accuracy, reproducibility, and scalability \cite{brady2022cataloguing, searle2024nanopore}.

Recent advancements in click chemistry and bioorthogonal chemistry \cite{stump2022click,koniev2015developments,scinto2021bioorthogonal} have attempted to address this issue by enabling the identification of specific amino acids and their positions. For instance, Zheng et al. demonstrated the sequencing of short antibody peptides using targeted amino acid labeling \cite{Zheng2024.05.31.596913}. However, these techniques are still limited by the number of amino acids that can be correctly identified, resulting in partially masked sequences (e.g., xCxxCxxx, where C is the experimentally identifiable amino acid) \cite{swaminathan2018highly}. Additionally, the click chemistry platform demonstrated in Zheng et al. only works with non-native peptides that have undergone \textit{a priori} chemical modifications\cite{Zheng2024.05.31.596913}; limitiations in this step means that parts of the proteomic retinue are not measurable with this approach. Our language model can work with input from this non-native peptide platform, as well as hypothetical future developments in bioorthogonal chemistry that will allow for edman degregation of native peptides. 

To address this specific limitation, we propose a novel approach leveraging pretrained language models. Large language models (LLMs) have shown remarkable adaptability in interpreting protein sequences, excelling in predicting structures, functions, and evolutionary relationships \cite{bepler2021learning, ruffolo2024designing, lv2024prollama}. We hypothesize that these models can be used to predict the identity of amino acids that are conditionally difficult to determine experimentally.

In this paper, we present a method that simulates partial sequencing data by selectively masking amino acids that are experimentally challenging to identify in protein sequences from the UniRef database. This targeted masking mimics real-world sequencing limitations. We then utilize ProtBert \cite{9477085}, a transformer-based model, to predict these masked residues, providing a probabilistic reconstruction of the complete protein sequence.

We evaluate our approach on three \textit{Escherichia} bacterial species: \textit{E. coli}, \textit{E. albertii}, and \textit{E. fergusonii}. Our results demonstrate high prediction accuracy even with extremely limited known amino acids.
We also validate the biological relevance of our predictions through structural assessment using AlphaFold \cite{jumper2021highly}, and standard structure evaluation metrics such as template modeling score (TM-score)\cite{xu2010significant,zhang2005tm} and the local distance difference test (lDDT)\cite{mariani2013lddt}.

This innovative integration of simulated experimental constraints with computational predictions offers a promising avenue for enhancing protein sequence analysis. By improving our ability to interpret partially sequenced data, we aim to accelerate advancements in proteomics and structural biology, potentially unlocking new insights into protein structure and function.

The remainder of this paper is structured as follows: In the Methods section, we detail our data preparation, model fine-tuning process, and evaluation metrics. The Results section presents our comprehensive analysis of our model's performance across various scenarios. Finally, we discuss the implications of our findings and potential future directions in the Discussion section.
\section{Problem statement}

In an assumption that the partial sequencing can be acquired from Edman degradation enhanced by click chemistry, which provides the positions and identities of a limited set of amino acids within a protein, we aim to predict the complete protein sequence. This task involves using an LLM modified and finetuned from BERT/ProtBERT \cite{devlin2018bert,10.1093/bioinformatics/btac020} to fill in the gaps from unknown amino acids, given the context provided from the known ones in combination with the protein's domain constraint, determined at the species level. Our goal is to develop a computational approach that can accurately predict the full protein sequence from this partial information, potentially revolutionizing protein sequencing methodologies.

\section{Methods}

\subsection{Protein Dataset}

\begin{figure}
  \centering
  \includegraphics[width=\columnwidth]{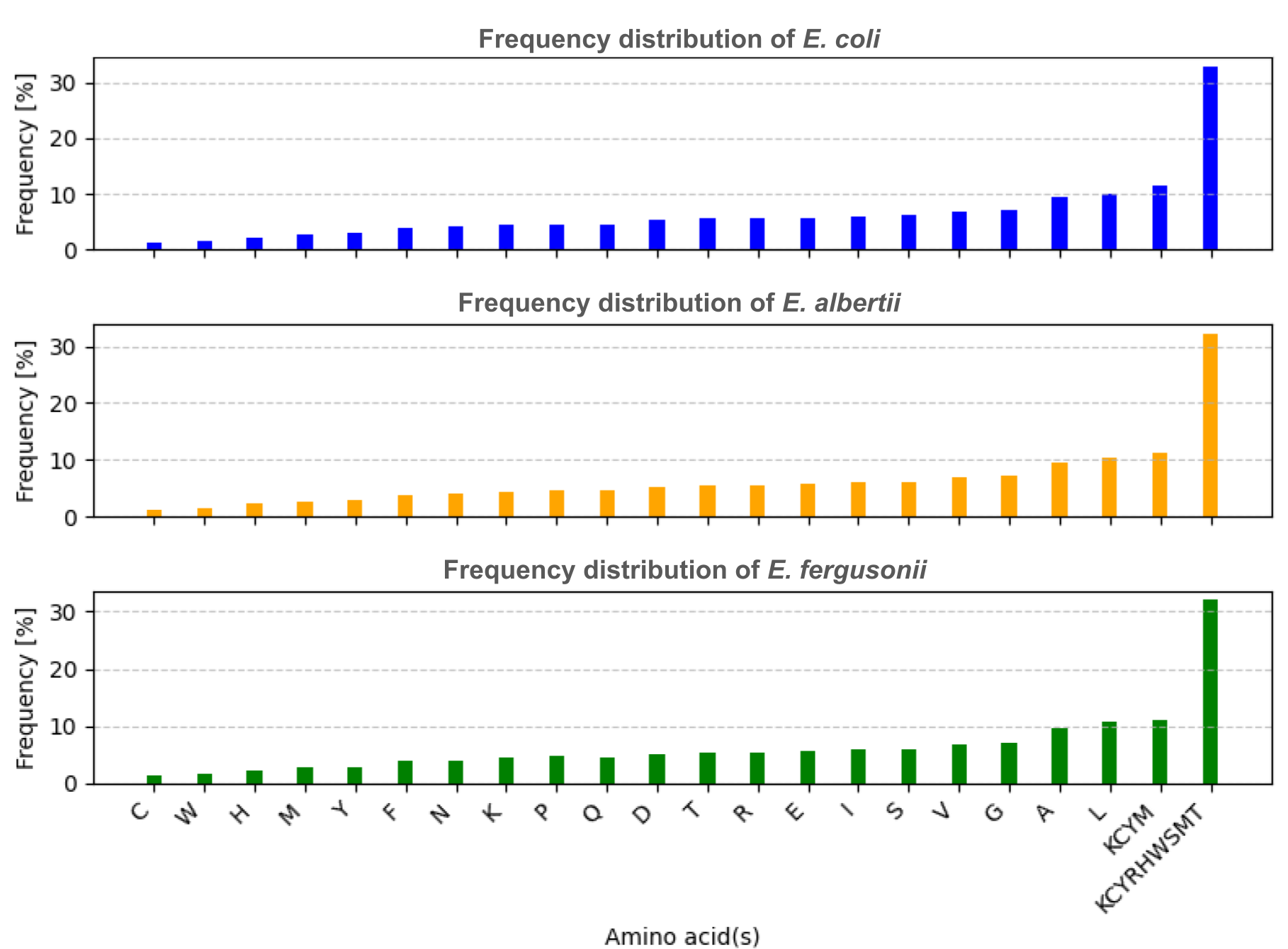}
  \caption{Frequency distribution across different amino acids and sets in \textit{E. coli} (top), \textit{E. albertii} (middle), and \textit{E. fergusonii} (bottom).}
  \label{fig:data-analysis}
  \Description{}
\end{figure}

For model training, which we conducted on 8 NVIDIA DGX A100 80GB cards, we utilized the UniProt Reference Clusters (UniRef) database\cite{10.1093/bioinformatics/btu739}, specifically UniRef100, focusing on three bacterial species: \textit{Escherichia coli} (NCBI taxID 652), \textit{Escherichia albertii} (NCBI taxID 208962), and \textit{Escherichia fergusonii} (NCBI taxID 564). The chosen dataset combined identical sequences and subfragments with 11 or more residues into one UniRef100 entry, reducing potential data leakage between training, evaluation, and testing datasets. Additionally, we removed sequences from hypothetical protein group to ensure the biological relevance of the dataset. Following the pretrained model's data processing, we mapped non--canonical or unresolved amino acids ([BOUZ]) to \textit{unknown} (X)\cite{9477085}. The frequency distribution of amino acids extracted from the three species is presented in Figure~\ref{fig:data-analysis}.

We propose working on two cases of targeted sets of amino acids. The first set (\textbf{KCYM}) contains amino acids with two or more publications supporting successful identification: Lysine (K)\cite{tantipanjaporn2023development, anderson1964use, anderson1963n}, Cysteine (C)\cite{vantourout2020serine, renault2018covalent, grant2017modification}, Tyrosine (Y)\cite{ban2013facile, ABDULFATTAH201887, szijj2020tyrosine, liu2019biocompatible}, and Methionine (M)\cite{lin2017redox, zang2019chemoselective}. The second set (\textbf{KCYMRHWST}) includes the amino acids from the first set, with additional amino acids that have at least one publication supporting successful identification: Arginine (R)\cite{wanigasekara2018arginine}, Histidine (H)\cite{wan2022histidine}, Tryptophan (W)\cite{decoene2022triazolinedione}, Serine (S)\cite{vantourout2020serine}, and Threonine (T)\cite{webster2014mild}.

\subsection{Training Model}

We chose ProtBERT\cite{10.1093/bioinformatics/btac020} as our pretrained model due to its well performance in general tasks, lightweight nature (420M parameters), and bidirectional property. However, we modified the architecture of the model to use a masked language modeling head for our training task, compatible with our problem formulation. We trained one model per domain (species) and per set of amino acids, resulting in a total of six finetuned and architecturally modified models. For \textit{E. coli} and \textit{E. albertii}, we performed training and evaluation on 50k and 25k sequences, respectively. Due to data limitation, \textit{E. fergusonii} was trained and evaluated on 40k and 5k sequences, respectively. Given the extremely high masking rate (67--88\%, see Figure~\ref{fig:data-analysis} and Table~\ref{table: generalizability}), we removed any totally--masked sequences before constructing the training and evaluation datasets. The pretrained and architecturally modified model was then finetuned using HuggingFace transformers\cite{wolf2019huggingface}.

\begin{figure*}
  \centering
  \includegraphics[width=\textwidth]{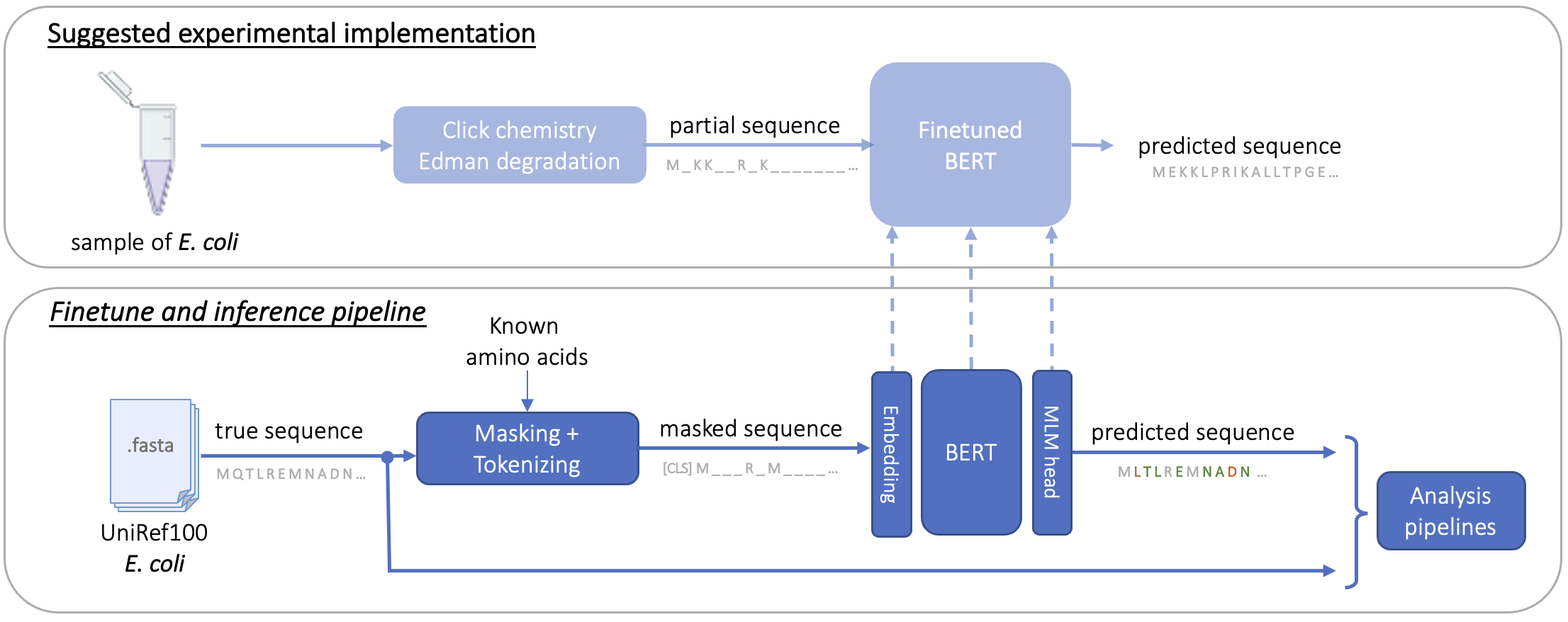}
  \caption{Overview of pipeline.}
  \label{fig:pipeline}
  \Description{}
\end{figure*}

During the training process, we followed the ProtBERT's pretrained tokenization scheme: one token per residue. Any residue not in the set of known amino acids was set to [MASK], and sequences were padded or truncated to a length of 1024. The training process used batch size of 50, and was evaluated using cross--entropy loss and unmasking accuracy. All models were trained on an A100 GPU and 16 CPUs. The overview of the training pipeline is visualized in Figure~\ref{fig:pipeline}.

\subsection{Evaluation Strategies}

\begin{figure}[t!]
  \centering
  \includegraphics[width=\columnwidth]{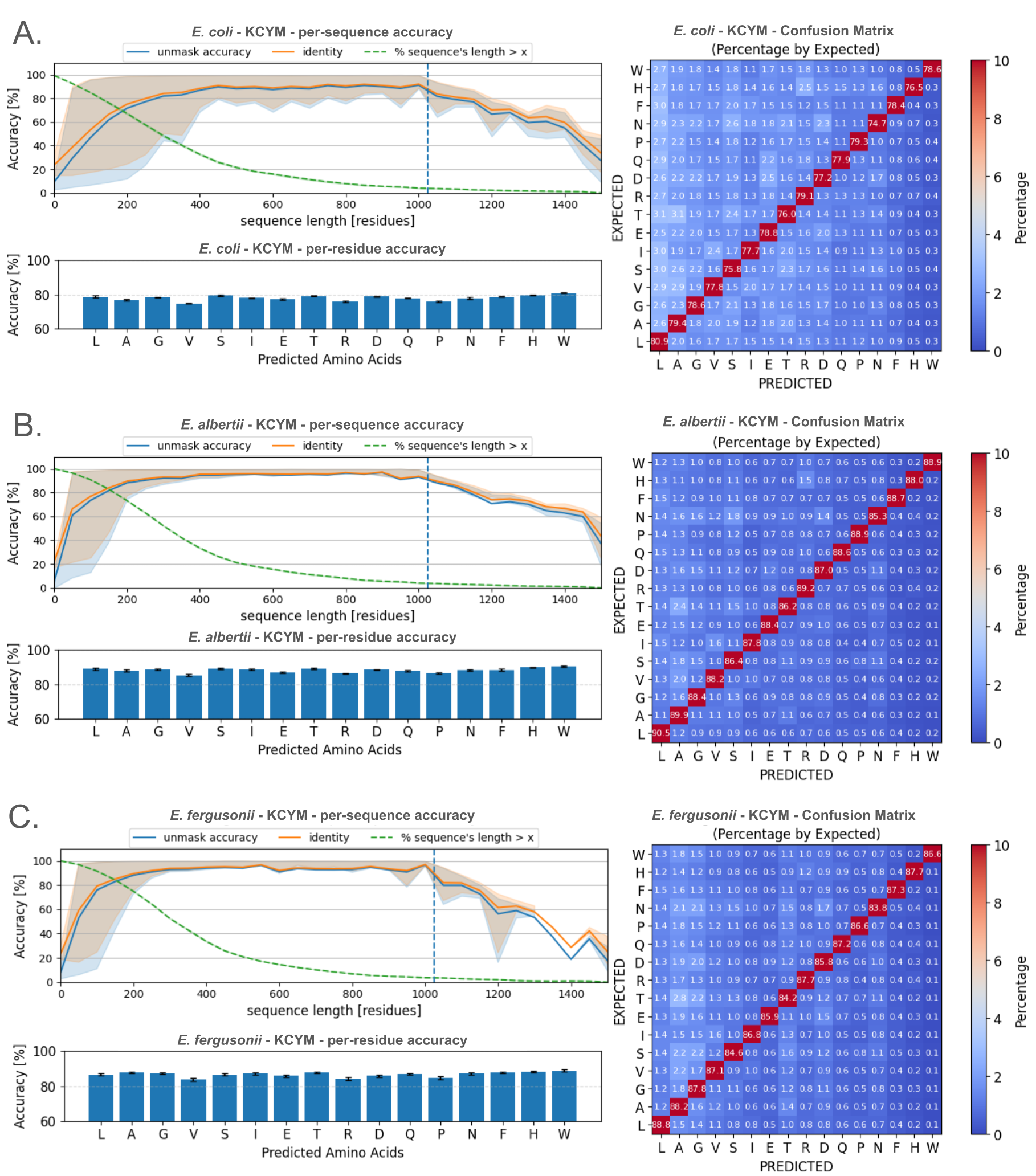}
  \caption{The accuracy per amino acid (right), and the per--sequence average accuracy of unmasking (top left) and identity (bottom left) of (A) \textit{E. coli} (B) \textit{E. albertii} and (C) \textit{E. fergusonii} with known amino acids: KCYM}
  \label{fig:accuracy-KCYM}
  \Description{}
\end{figure}

The performance of the model predictions were evaluated based on two major aspects: prediction accuracy and secondary structure similarity. For prediction accuracy, we computed three measures to compare the primary sequence of the predicted and the true proteins: per--token accuracy, average per--sequence unmasking accuracy (i.e, excluded known amino acids), and average per--sequence total identity. Beside using an in--domain inference dataset to study the performance of models (see Figure~\ref{fig:accuracy-KCYM} and \ref{fig:acuracy-KCYMRHWST}), we also examined cross--domain accuracy among the three species. This aims to observe how taxonomic metrics (\textit{a prior knowledge about evolutionary distance in a phylogenic tree} correlates with the performance of our model predictions (see Table~\ref{table: Evolution}). For the two sets of amino acids (\textbf{KCYM} and \textbf{KCYMRHWST}), we performed testing inference on 5,000 sequences per species (randomly sampled for 3 folds).

To present useful amino acid suggestions/prioritization for experimental development in click chemistry based amino acid identification in the wet lab, we also performed training with amino acids from the small set and one additional amino acid from ([RHWST]), creating five additional study cases: \textbf{KCYMR}, \textbf{KCYMH}, \textbf{KCYMW}, \textbf{KCYMS}, and \textbf{KCYMT}. This configuration was only applied to the \textit{E. coli} domain (with the same training protocols), and the inference was done using 25k sequences (one fold) of \textit{E. coli} (see Table~\ref{table: generalizability}).

\begin{figure*}
  \centering
  \includegraphics[width=\textwidth]{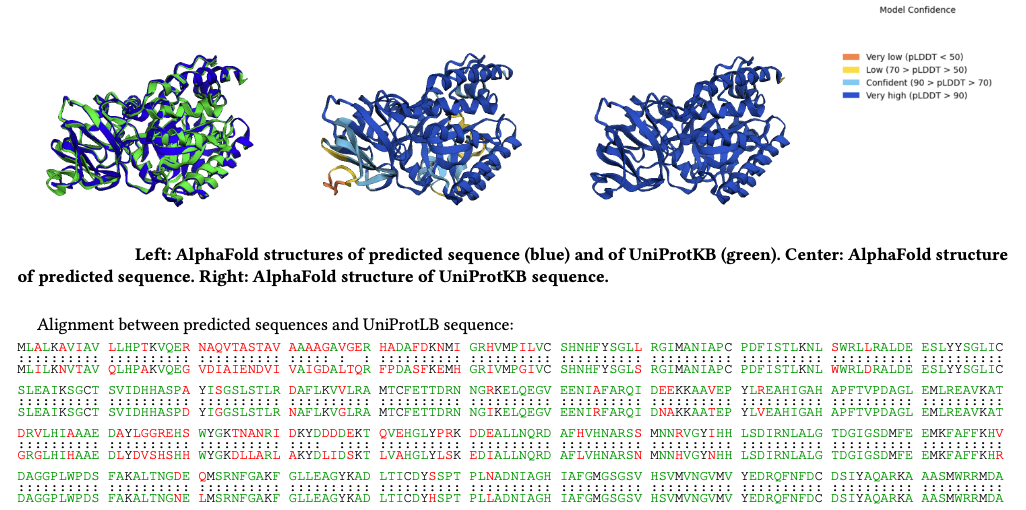}
  \caption{An illustrated example of structure based model validations.}
  \label{fig:alphafold}
  \Description{}
\end{figure*}

For the second aspect of measuring the quality of our predictions, we analyzed an important property of proteins: structure. AlphaFold\cite{jumper2021highly} is renowned for its high--accuracy prediction of protein three--dimensional structures from amino acid sequences using multiple sequence alignment in combination with a deep learning architecture. Recently, these AlphaFold predicted structures have been widely adapted inside large annotated databases, such as UniProt KnowledgeBase (UniProtKB)\cite{uniprot2023uniprot}. In this study, we used the AlphaFold platform to examine how predicted sequences with less than 90\% unmasking accuracy impact their structural integrity. Our study centered on sequences from the \textit{E. coli} inference (fold--1), derived from the \textbf{KCYM} case, with unmasking accuracy bounded to the range [50--90]\%. We used the reduced database in AlphaFold settings to generate structure predictions for our unmasked sequences, and the available AlphaFold structure of the true sequences (only those annotated in the UniProtKB). Filtering under these criteria yielded a total of 124 sequences for our structure analysis (see Result~\ref{section: structure}). Figure~\ref{fig:alphafold} visualizes the protein structure derived from the predicted sequence and the actual UniProtKB sequence, these two structures overlaid, as well as the alignment between the predicted and actual amino acid sequence for one of these 124 proteins (PDB A0A7H9QJ10). 

We computed the TM--score\cite{xu2010significant,zhang2005tm} to compare the global similarity between the topologies of two structures. For local similarity, we computed the local difference distance test of the backbone atoms (lDDT--C$\alpha$)\cite{mariani2013lddt,biasini2013openstructure} between the two structures, similar to the AlphaFold paper (see Figure~\ref{fig:structure-score}).

\section{Results}

\subsection{Inference Accuracy}
\label{section: inference accuracy}

\begin{figure}[t!]
  \centering
  \includegraphics[width=\columnwidth]{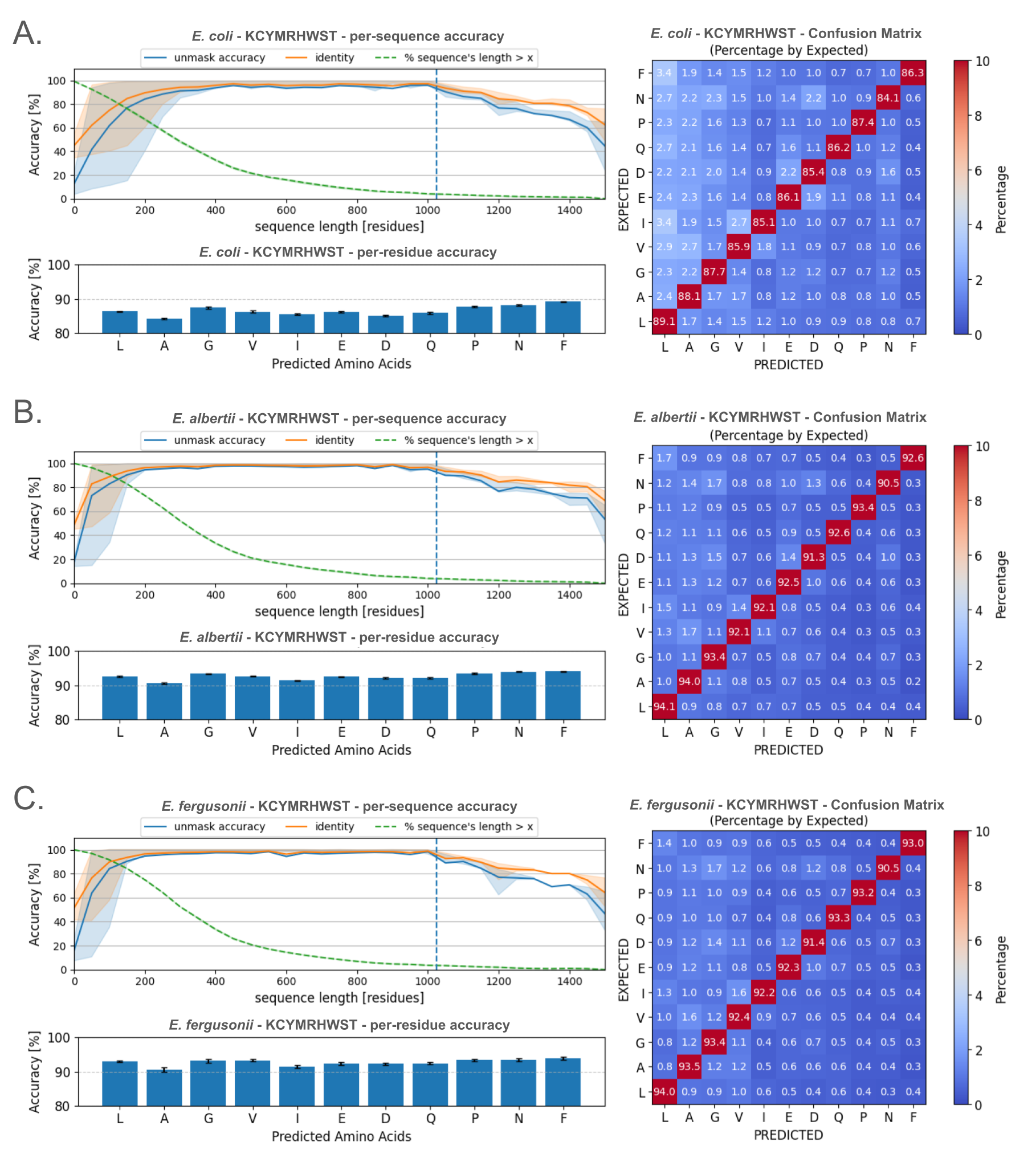}
  \caption{The accuracy per amino acid (right), and the per--sequence average accuracy of unmasking (top left) and identity (bottom left) of (A) \textit{E. coli} (B) \textit{E. albertii} and (C) \textit{E. fergusonii} with known amino acids: KCYMTHWST}
  \label{fig:acuracy-KCYMRHWST}
  \Description{}
\end{figure}

The accuracy of sequence predictions generated using the known amino acids set \textbf{KCYM} is presented in Figure~\ref{fig:accuracy-KCYM}, and the set \textbf{KCYMRHWST} is presented in Figure~\ref{fig:acuracy-KCYMRHWST}. As shown in the confusion heatmap for \textbf{KCYM}, even with a masking rate of 88.5\%, the per--amino--acid accuracy reaches 74.7--80.9\% in \textit{E. coli}, 85.3--90.5\% in \textit{E. albertii}, and 83.8--88.8\% in \textit{E. fergusonii}. 

The top left panel of Figure~\ref{fig:accuracy-KCYM} shows that the average per--sequence accuracies (unmasking and identity) vs. sequence length, averaged per 50-residue bin and highlighted using the 75th percentile interval. The average per--sequence unmasking accuracy and identity are 73.53\% and 76.75\% for \textit{E. coli}, 88.46\% and 89.87\% for \textit{E. albertii}, 88.33\% and 89.73\% for \textit{E. fergusonii}. The performance of the model decayed when the protein sequence length exceed the model's maximum length of 1024 residues. This behavior is expected due to the property of the BERT model, which has linear positional embedding and the training maximum length is set to be 1024 residues. Note that only about 5\% of sequences in the data had length exceeding this threshold.

After taking this into account, the line plots (left) indicate that the performance of the prediction is more stable and accurate for longer protein sequences. Specifically, with just the four known amino acids \textbf{KCYM}, the unmasking accuracy for sequences longer than 300 residues reached approximately 80\% for \textit{E. coli} and over 90\% for \textit{E. albertii} and \textit{E. fergusonii}. However, it should be noted that only about half of the protein sequences in these species are longer than 300 residues.

In the case of knowing nine amino acids \textbf{KCYMRHWST} (see Figure~\ref{fig:acuracy-KCYMRHWST}), where the masking rate is 67.1\%, the per--amino--acid accuracy reaches 84.1--89.1\% in \textit{E. coli}, 90.5--94.1\% in \textit{E. albertii}, and 90.5--94.0\% in \textit{E. fergusonii}. The average per--sequence unmasking accuracy and identity are 83.26\% and 88.96\% in \textit{E. coli}, 93.38\% and 95.62\% in \textit{E. albertii}, 93.49\% and 95.69\% in \textit{E. fergusonii}. For proteins with length longer than 200 residues, representing 80\% of protein sequences, the unmasking accuracy of \textit{E. coli} exceeds 80\%, while \textit{E. albertti} and \textit{E. fergusonii} exceed 90\% accuracy.

\subsection{Evolution Inference}
\label{section: evolution}

We evaluated the model's performance in capturing evolutionary information by cross--inferring each species’ protein sequences using models trained on each other species. Our three species: \textit{E. coli}, \textit{E. albertii}, and \textit{E. fergusonii}, are all members of the \textit{Escherichia} genus, and thus are expected to share a significant amount of genetic information, indicating a decent homology in protein sequences.

\begin{table*}[tb!]
\caption{Inference unmasking accuracy and standard deviation across species and sets of amino acids}
\centering
\resizebox{\textwidth}{!}{\begin{tabular}{|>{\centering\arraybackslash}p{3.5cm}|>{\centering\arraybackslash}p{3.2cm}|>{\centering\arraybackslash}p{3.2cm}|>{\centering\arraybackslash}p{3.2cm}|}
\hline
\multicolumn{4}{|c|}{\textbf{finetuning model on known tokens: KCYM (3 folds)}} \\ \hline
& \textbf{\textit{E. coli} model} & \textbf{\textit{E. albertii} model} & \textbf{\textit{E. fergusonii} model} \\ \hline
\textbf{\textit{E. coli}} & \textbf{73.53\% (0.49)} & 51.38\% (0.17) & 45.22\% (0.09) \\ \hline
\textbf{\textit{E. albertii}} & 65.15\% (0.49) & \textbf{88.46\% (0.18)} & 50.16\% (0.64) \\ \hline
\textbf{\textit{E. fergusonii}} & 60.06\% (0.35) & 50.61\% (0.34) & \textbf{88.33\% (0.15)} \\ \hline
\hline
\multicolumn{4}{|c|}{\textbf{finetuning model on known tokens: KCYMRHWST (3 folds)}} \\ \hline
& \textbf{\textit{E. coli} model} & \textbf{\textit{E. albertii} model} & \textbf{\textit{E. fergusonii} model} \\ \hline
\textbf{\textit{E. coli}} & \textbf{83.26\% (0.45)} & 69.08\% (0.33) & 64.02\% (0.17) \\ \hline
\textbf{\textit{E. albertii}} & 82.35\% (0.45) & \textbf{93.38\% (0.05)} & 71.04\% (0.67) \\ \hline
\textbf{\textit{E. fergusonii}} & 79.58\% (0.46) & 72.49\% (0.18) & \textbf{93.49\% (0.13)} \\ \hline
\end{tabular}}
\label{table: Evolution}
\end{table*}

As shown in Table~\ref{table: Evolution}, when the model only knows the four amino acids \textbf{KCYM}, the unmasking accuracy is high only when the training and inference are from the same domain (see Result \ref{section: inference accuracy}). The unmasking accuracy is significantly lower when the model tries predicting out--of--domain sequences. The results of the \textbf{KCYM} case reveal that, in the condition where the domain is specified, the model predictions only need a small set of known amino acids (in this case, \textbf{KCYM}) to capture the characteristics of the domain's protein sequences, achieving an average accuracy of at least ~73\%. However, with this size of amino acids set, our model fails to capture the nuance of sequences beyond the domain specified.

In the case of knowing nine amino acids \textbf{KCYMRHWST}, the in--domain unmasking accuracy increased by 5--10\% compared to the previous \textbf{KCYM} case. Besides the high in--domain accuracy, the model predictions for out--of--domain sequences also performed much better, with the lowest accuracy at 64.02\% when training on \textit{E. fergusonii} and inferring on \textit{E. coli}, and the highest accuracy at 82.35\% when training on \textit{E. coli} and inferring on \textit{E. albertii}. 

Overall, the model trained on \textit{E. coli} performed best on out--of--domain inference, followed by \textit{E. albertii}, and lastly \textit{E. fergusonii}. This outcome is expected due to the high yield of protein sequences available from \textit{E. coli} and \textit{E. albertii} compared to \textit{E. fergusonii}. In summary, the cross--inference results indicate that knowledge of the species to which a sequence belongs increases prediction accuracy. Furthermore, they demonstrate the model’s potential for predicting protein sequences based on another related species when the sequences’ species identity may not be known.

\subsection{Generalizability}
\label{section: generalizability}

\begin{table*}[t!]
\caption{Inference accuracy report and masking ratio of different sets of known amino acids}
\centering
\resizebox{\textwidth}{!}{\begin{tabular}{|c|c||c|c|c|c|c||c|}
\hline
& \textbf{KCYM} & \textbf{KCYMR} & \textbf{KCYMH} & \textbf{KCYMW} & \textbf{KCYMS} & \textbf{KCYMT} & \textbf{KCYMRHWST} \\ \hline
\textbf{Per--token acc. [\%]} & 78.17\% & 80.26\% & 78.85\% & 79.52\% & 78.70\% & \textbf{80.38\%} & 86.82\% \\ \hline
\textbf{Per--seq acc. [\%]} & 73.53\% & \textbf{76.28\% (32.26)} & 75.00\% (33.25) & 75.83\% (32.92) & 75.22\% (32.91) & 75.74\% (32.53) & 83.26\% \\ \hline
\textbf{Per--seq identity [\%]} & 76.75\% & \textbf{80.55\% (26.41)} & 78.57\% (28.46) & 79.12\% (28.40) & 79.83\% (26.71) & 80.08\% (26.67) & 88.96\% \\ \hline \hline
\textbf{Masking ratio [\%]} & 88.47\% & 82.78\% (3.73) & 86.30\% (3.40) & 86.92\% (3.28) & \textbf{82.34\% (3.69)} & 82.88\% (3.43) & 67.09\% \\ \hline
\end{tabular}}
\label{table: generalizability}
\end{table*}

From previous results (see Results~\ref{section: inference accuracy} and \ref{section: evolution}), our work suggests that transformer models like BERT can predict protein sequences with high accuracy, given prior knowledge of limited sets of amino acids and the species domain. Additionally, the accuracy of the predictions increases significantly with a larger set of known amino acids. However, expanding the set of identifiable amino acids introduces exponential challenges in Edman degradation. This process requires peptides to undergo more chemical identification cycles, leading to an increased noise in sequencing and a higher risk of unstable peptide degradation. Therefore, we investigated our model performance on sequence prediction by using five different sets of known amino acids as a guide for prioritizing amino acids to develop future click chemistry based identification for; in essence we are comparing how unmasking new amino acids ameliorate model performance, and comparing these results to priortize future wet lab efforts. We evaluated the inference of five additional models, which are trained on five known amino acids: four being \textbf{KCYM} and one from the set ([RHWST]) amino acids (see Table~\ref{table: generalizability}).

Among the five experiments, the one with \textbf{KCYMS} has the highest sequence coverage from known amino acids, at 17.66\% (corresponding to 82.34\% masking rate). However, the case of \textbf{KCYMR} demonstrates the best, with average per--sequence unmasking accuracy at 76.28\% (2.75\% more than \textbf{KCYM}), average per--sequence identity at 80.55\% (3.8\% more), and per--token accuracy at 80.26\% (2.09\% more). The other four cases show comparable results to \textbf{KCYMR} (within ~2\% differences).

\subsection{Structure Analysis}
\label{section: structure}

\begin{figure}
  \centering
  \includegraphics[width=\columnwidth]{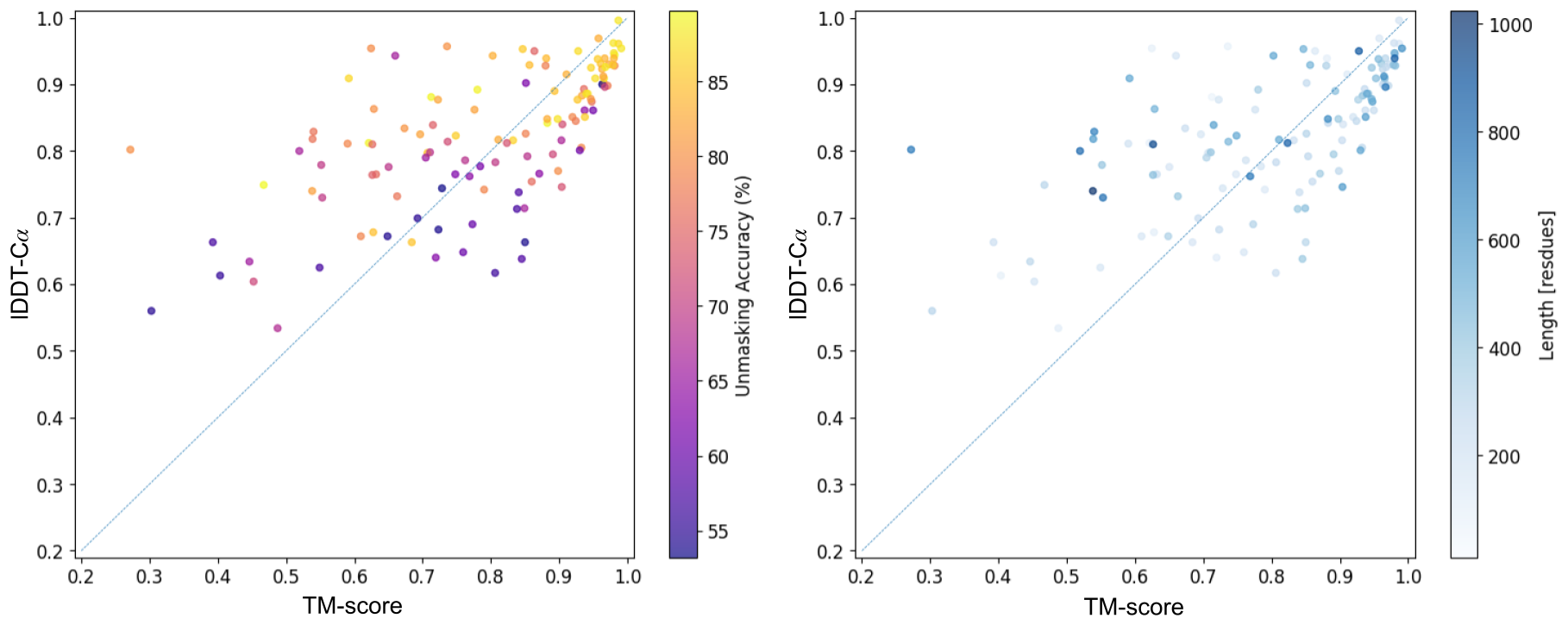}
  \caption{ C$\alpha$ LDDT vs TM--score generated from comparing \textit{E. coli} -- KCYM's predicted vs real sequence's AlphaFold structures, colored in unmasking accuracy (left) and sequence length (right)}
  \label{fig:structure-score}
  \Description{}
\end{figure}

The comparison of lDDT--C$\alpha$ vs. TM--score between the predicted and true sequences' AlphaFold structures is shown in Figure~\ref{fig:structure-score}, in which the left panel is colored by unmasking accuracy, and the right panel is colored by sequence length. 

The TM--score evaluates the global similarity between two structures, while the lDDT--C$\alpha$ assesses the local distances of the backbones. According to the plot, the high lDDT--C$\alpha$ can happens even with low TM--score, but not the opposite where the TM--score is high but lDDT--C$\alpha$ is low. This is often caused by the structures having the large difference in the bending angles at the coil regions, yield a divergence in the structure's global shapes, and hence resulted in low TM--score. But the local structure conformations, such as alpha--helices and beta--sheets, are conserved, leading to high value of lDDT--C$\alpha$. An AlphaFold example result (UniProtKB ID: A0A066T1W5) is presented in Figure~\ref{fig:alphafold}, showing the molecular view of two structures (using py3Dmol\cite{py3Dmol}) and their pairwise alignment, colored by unmasking matches (green) and mismatches (red). An additional illustrative example of a different protein is presented in the appendix.

To know how our model prediction's quality (measured by unmasking accuracy) impacts the predicting protein structure, we need to understand how the value of TM--score approximately corresponds to whether the protein pairs sharing the same topology. Xu et al.'s paper, studied on the CATH and SCOP databases, reported that the high posterior probability of two structures having the same topology corresponds to a TM--score roughly between 0.4 and 0.6, with the specific threshold varies by datasets\cite{xu2010significant}. In our structure results, reported from 124 samples of \textit{E. coli} with known set as \textbf{KCYM}, we also observed the decrease in general lDDT--C$\alpha$ values when the TM--score lower than 0.6, and hence 0.6 is our evaluation threshold. This means for TM--score > 0.6, we have a high statistical confidence that the two structures are the same topology. And for TM--score < 0.6, we need to evaluate auxiliary metrics such as lDDT--C$\alpha$, unmasking accuracy, sequence length, etc. to conclude the similarity in topology.

The Figure~\ref{fig:structure-score}'s left panel suggests that we have a high confidence in structure similarity between our predicted sequences and true sequences (TM--score > 0.6) when the unmasking accuracy is above ~75\%. For outliers where the unmasking accuracy > 75\% but TM--score < 0.6, we notices that their sequence length are often long (see Figure~\ref{fig:structure-score}'s right panel). Because of the sequence length, these protein are thus expected to have higher chance having local divergence, leading to a sensitive TM--score, but the high lDDT--C$\alpha$. 

However, although the low accuracy predictions are thought to has low structure similarity, it is not the case. For TM--core > 0.6, many of the sequences has unmasking accuracy < 75\%, and some are even less than 65\%. These sequences are observed to often have lower lDDT--C$\alpha$ compared to ones with high accuracy.

\section{Future Directions}

Peptide sequencing enabled by our language model will have many important implications for the development of liquid biopsies, which could yield more information for treatment decisions in the oncology clinic. Liquid biopsy is a minimally invasive tool to identify cancer biomarkers within fluids such as blood plasma and urine. These liquid samples have been readily explored as sources of nucleotide biomarkers such as non-coding RNAs and tumor-specific DNA, but creating diagnostics based on proteins has been limited by signal to noise ratios for the detection of low-abundance hits, and difficulty discerning the source of proteins to cancer- specific cells without first isolating the circulating cancer cells \cite{marchioni2021biomarkers}. However, liquid biopsies are advantageous due to their safety, high repeatability, ability to monitor disease progression and prognosis, all without the need for an inpatient procedure \cite{bergerot2018role}.

To this end, recent advances have been focused on the annotation of the cell-free transcriptome of plasma \cite{koh2014noninvasive}\cite{vorperian2022cell} and urine \cite{vorperian2023multiomics}\cite{lin2017emerging}\cite{sin2017deep} to identify candidate RNAs for disease progression. 
There are notable protein signatures that are highly informative related to disease progression and cancer treatment prognosis which are present for clear-cell renal cell carcinomas (ccRCC) \cite{marchioni2021biomarkers}, prostate cancer\cite{massoner2014epcam}, and urothelial cancers \cite{dressler2024proteomic}\cite{dressler2023epcam}. These changes are notable for multiple proteoforms: immune checkpoint proteins PD-1/PDL-1 \cite{gulati2021biomarkers}\cite{larrinaga2021soluble}, CTLA4 \cite{gulati2021biomarkers}, epithelial cell adhesion molecule (EpCAM), and glycosylation changes in the protein EpEX \cite{dressler2023epcam}\cite{fellinger2008glycosylation}. These changes are currently monitored by immuno- histochemical (IHC) assessments of tumor biopsies. IHC is difficult to apply widely to novel biomarker discovery and is limited to the efficacy of antibodies employed in these assays. Notably, unless antibodies are specifically generated for cancer-specific proteoforms, getting information on these proteoforms is an arduous task. 

Thus, there is a significant technological gap between current diagnostic assays and the proteoform resolution necessary to characterize and quantifiably identify cancer-specific biomarkers and prognosis indicators \cite{di2020searching}. Taken together, an improvement in proteoform identification and quantification with resolution to the single molecule would foster rapid development and implementation of utilizing well- studied proteoforms as both diagnostic and prognosis biomarkers. Optimally, such a technology would enable the detection of protein analytes in fecal, urine, or plasma samples.

Urine samples are safe and relatively simple to work with, have minimal variation in sample complexity compared to blood plasma, and contain signatures of genitourinary cells. We have identified urine as a “gold standard” for a liquid biopsy that is facile to collect and prepare for analysis that will likely contain invaluable information related to cancer diagnosis and prognosis. Thus, in the future, we aim to expand peptide sequencing via the language model presented in this paper to develop a platform to directly sequence proteins within a complex milieu through highly-specific chemical ligation of amplifiable DNA barcodes for amino acid identity, sequence position, and peptide identity which provides a quantifiable readout with higher sensitivity than mass spectrometry alone \cite{timp2020beyond}\cite{alfaro2021emerging}. This future platform will driven by closely entwined advancements in machine learning algorithm design and chemical reaction development and characterization.

\section{Discussion}

We present a protein language model designed to determine the complete sequence of a peptide based on the measurement of a limited set of amino acids. Traditional protein sequencing primarily relies on mass spectrometry, with some novel Edman degradation-based platforms capable of sequencing non-native peptides. However, these techniques face significant limitations in accurately identifying all amino acids, thus hindering comprehensive proteome analysis. Our approach simulates partial sequencing data by selectively masking amino acids that are experimentally challenging to identify in protein sequences from the UniRef database, thereby mimicking real-world sequencing limitations. By modifying and fine-tuning a ProtBert-derived transformer-based model, we predict these masked residues, providing an approximation of the complete sequence. Unlike traditional multiple sequence alignment (MSA) approaches, our model views sequence data as partial sequences, providing a new perspective and methodology for protein sequence analysis.

Our method, evaluated on three bacterial \textit{Escherichia} species, achieves per-amino-acid accuracy of up to 90.5\% when only four amino acids ([KCYM]) are known. Structural assessments using AlphaFold and TM-score validate the biological relevance of our predictions, and the model demonstrates potential for evolutionary analysis through cross-species performance. This integration of simulated experimental constraints with computational predictions offers a promising avenue for enhancing protein sequence analysis. By improving our ability to interpret partially sequenced data, our approach has the potential to accelerate advancements in proteomics and structural biology, enabling a probabilistic reconstruction of complete protein sequences from limited experimental data. 

Oxford Nanopore's DNA/RNA sequencing platform, which makes inferences from incomplete signal (squiggles) and converts them algorithmically to sequence space, initially performed with lower accuracy when first introduced \cite{brown2016nanopore, aver2015assessing} than our model in terms of accuracy of inferred sequence. This suggests that our computational model paired with a few additional wet lab improvements has the potential to yield the first clinically useful protein sequencing platform. 

However, several challenges remain. Experimental verification is essential to validate our computational predictions and ensure their biological relevance. Additionally, successfully implementing the hypothetical Edman degradation pipeline requires effective peptide immobilization techniques without C-terminus modification. Overcoming these hurdles will be crucial for the practical application of our method.

This integration of simulated experimental constraints with computational predictions offers a promising avenue for enhancing protein sequence analysis. By improving our ability to interpret partially sequenced data, we aim to accelerate advancements in proteomics and structural biology, potentially unlocking new insights into protein structure and function.

The future directions for our research involve expanding peptide sequencing via our language model to develop a platform capable of directly sequencing proteins within complex milieus. This will involve highly specific chemical ligation of amplifiable DNA barcodes for amino acid identity, sequence position, and peptide identity, providing a quantifiable readout with higher sensitivity than mass spectrometry alone \cite{timp2020beyond}\cite{alfaro2021emerging}. Such advancements will drive closely intertwined developments in machine learning algorithm design and chemical reaction characterization, ultimately fostering rapid implementation of proteoform-based diagnostics and prognostics.

In summary, our computational approach, validated through structural assessment using AlphaFold and TM-score, demonstrates significant potential for improving protein sequence analysis. By integrating these predictions with experimental techniques, we aim to bridge the gap between current technologies and the high-resolution identification required for advanced proteomics.

\begin{acks}
This section will be written after the blind review.
\end{acks}

\section{Code Availability}
Please use the following URL to access anonymized code during the review process: \url{https://github.com/aauthors131/protein-sequencing-LLMs.git}

\bibliographystyle{ACM-Reference-Format}
\bibliography{citations}

\onecolumn
\appendix
\section{Appendix}
\renewcommand{\figurename}{Supplement}
\setcounter{figure}{0} 

\begin{figure*}[ht!]
  \centering
  \subfigure[\textit{E. coli}]{\includegraphics[width=0.75\textwidth]{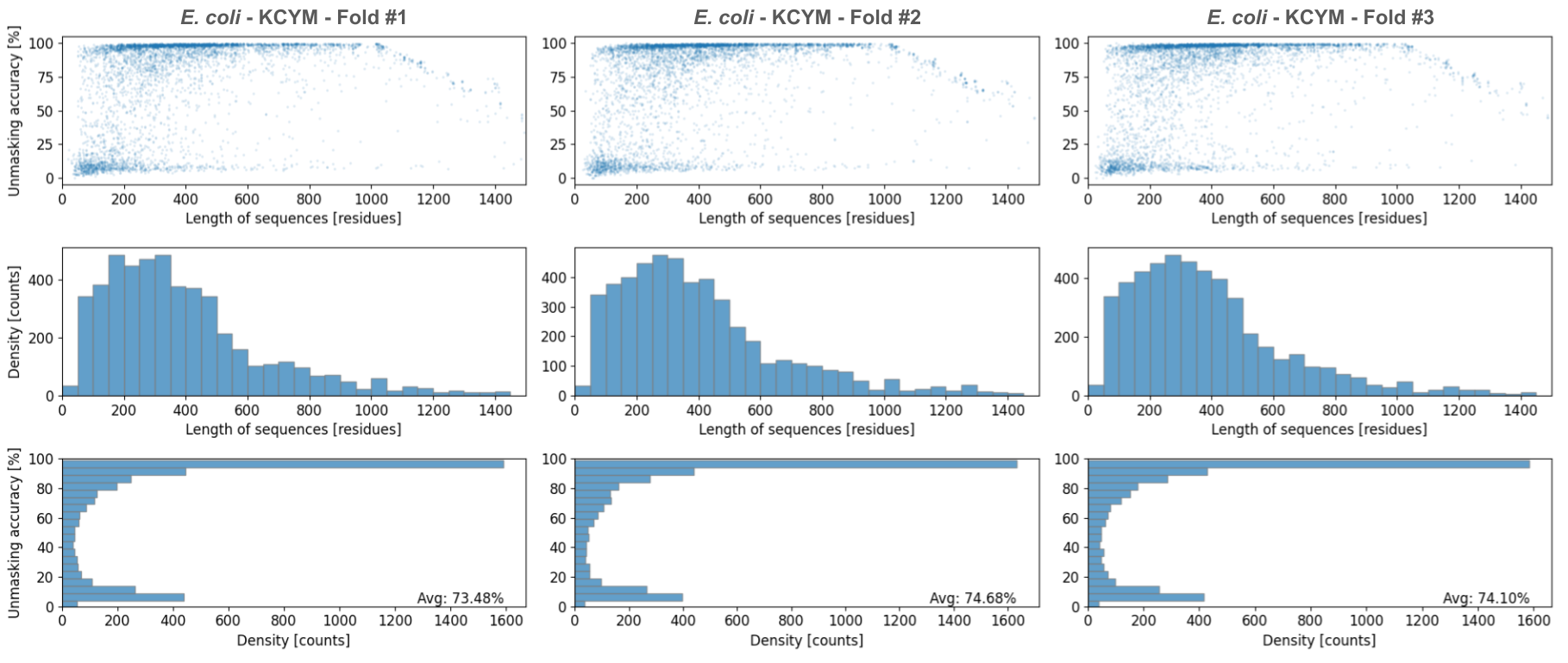}}
  \subfigure[\textit{E. albertii}]{\includegraphics[width=0.75\textwidth]{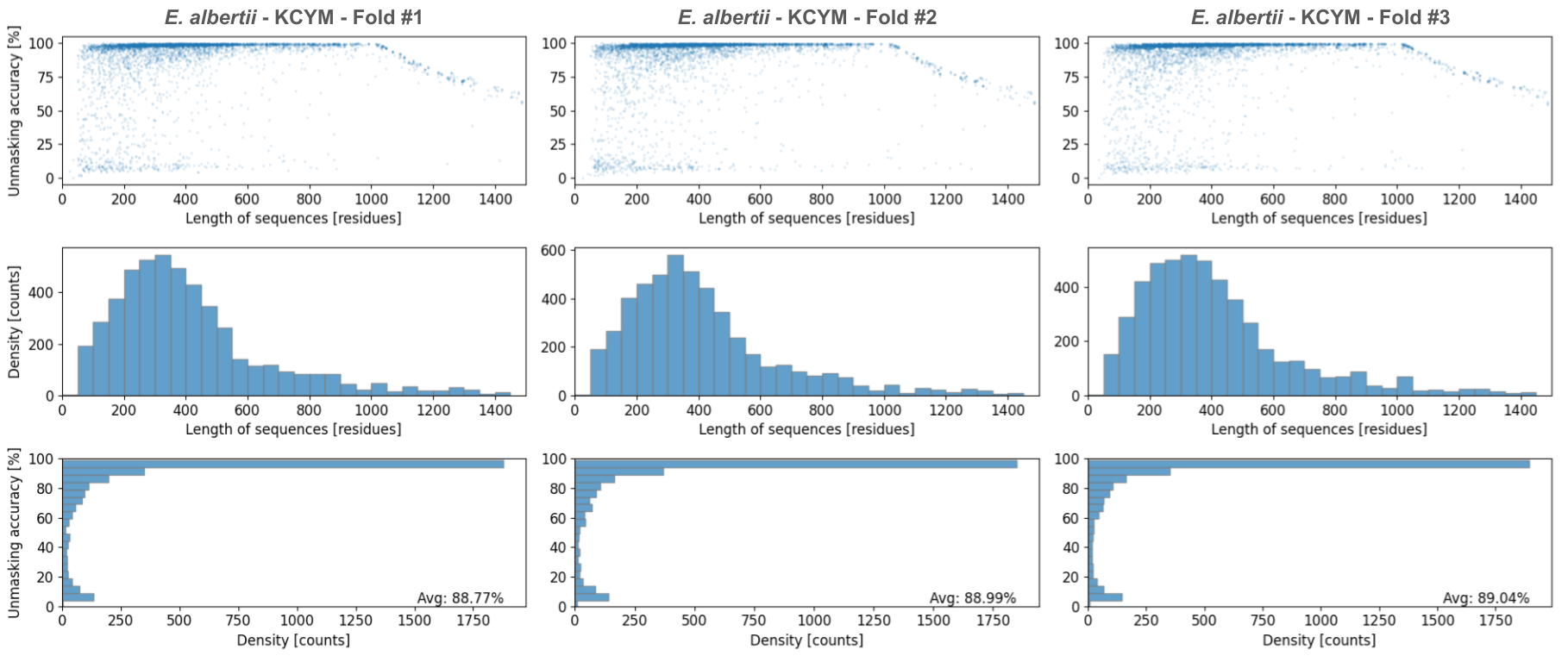}}
  \subfigure[\textit{E. fergusonii}]{\includegraphics[width=0.75\textwidth]{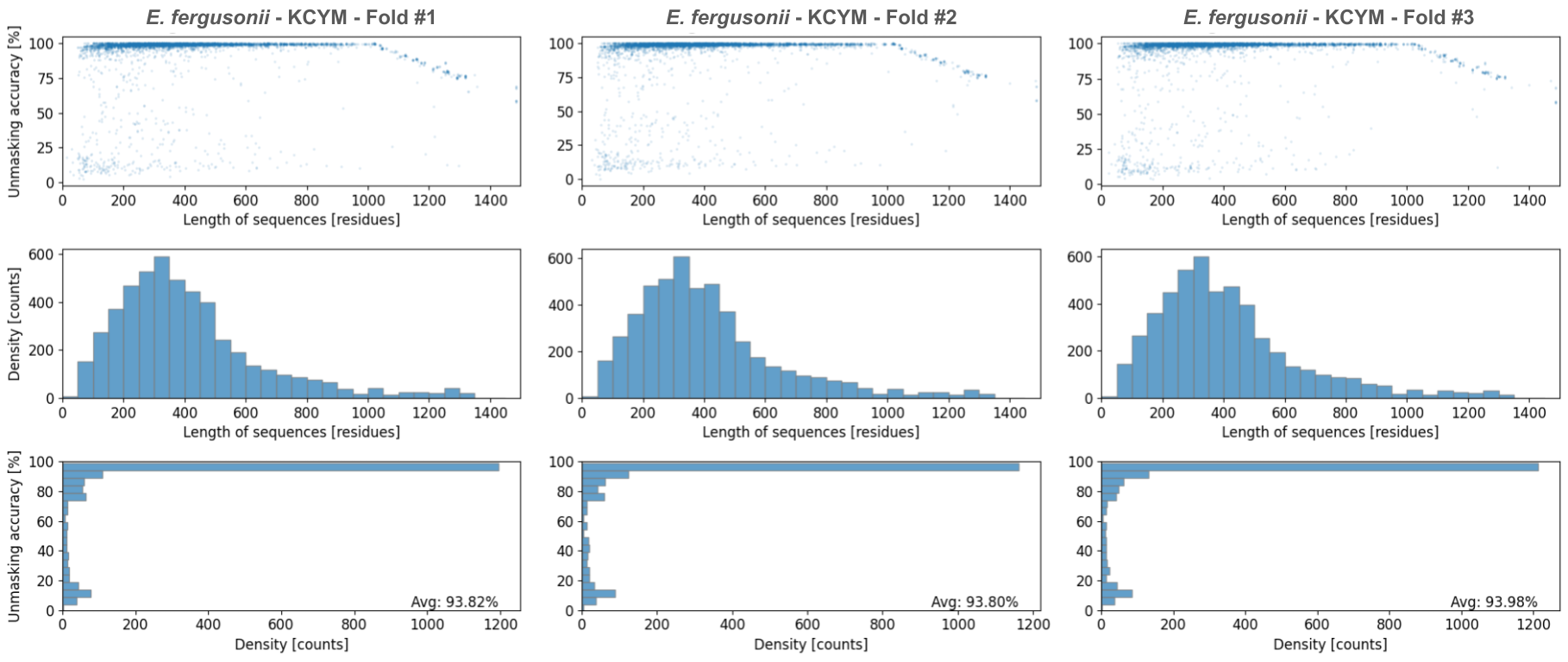}}
  \caption{The 3-fold unmasking inference results with density histograms from (a) \textit{E. coli}, (b) \textit{E. albertii}, and (c) \textit{E. fergusonii} with 4 known amino acids KCYM.}
  \label{sup:per-fold-KCYM}
  \Description{}
\end{figure*}

\begin{figure*}[ht!]
  \centering
  \subfigure[\textit{E. coli}]{\includegraphics[width=0.75\textwidth]{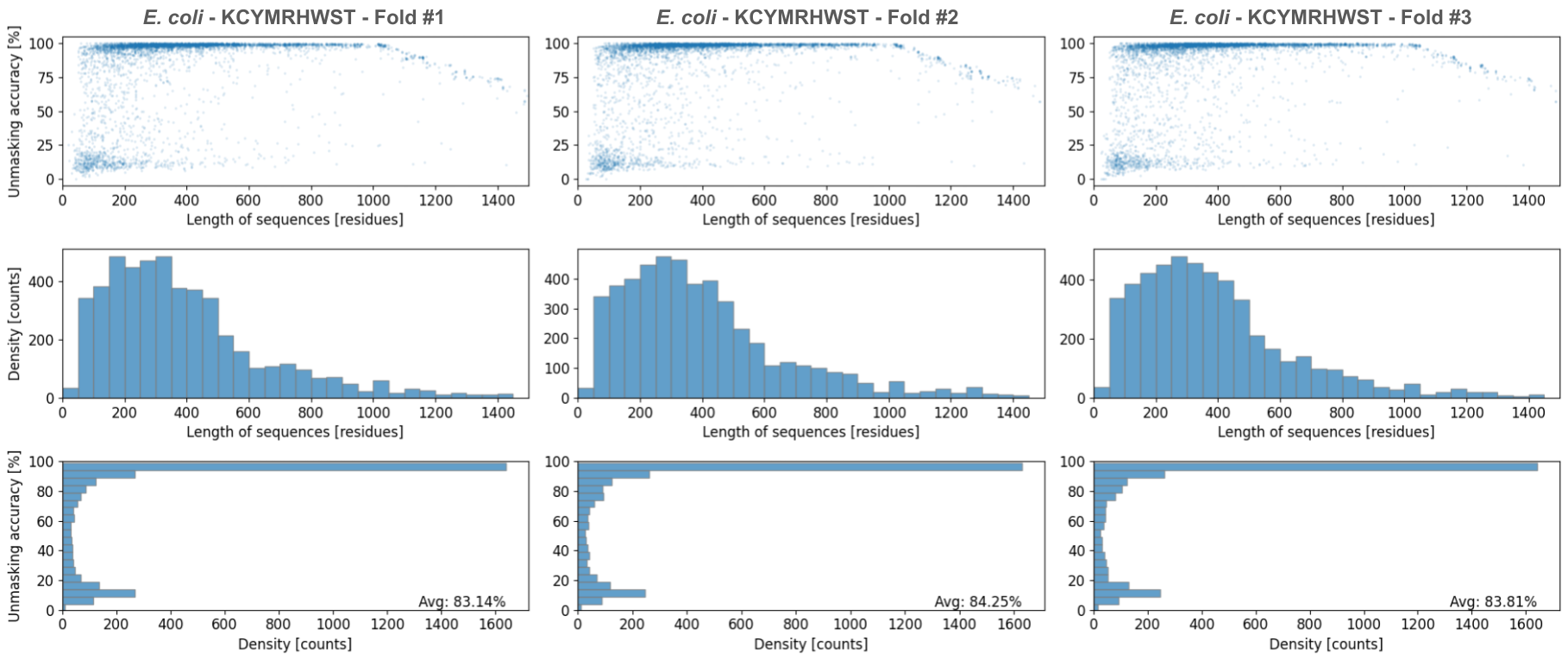}}
  \subfigure[\textit{E. albertii}]{\includegraphics[width=0.75\textwidth]{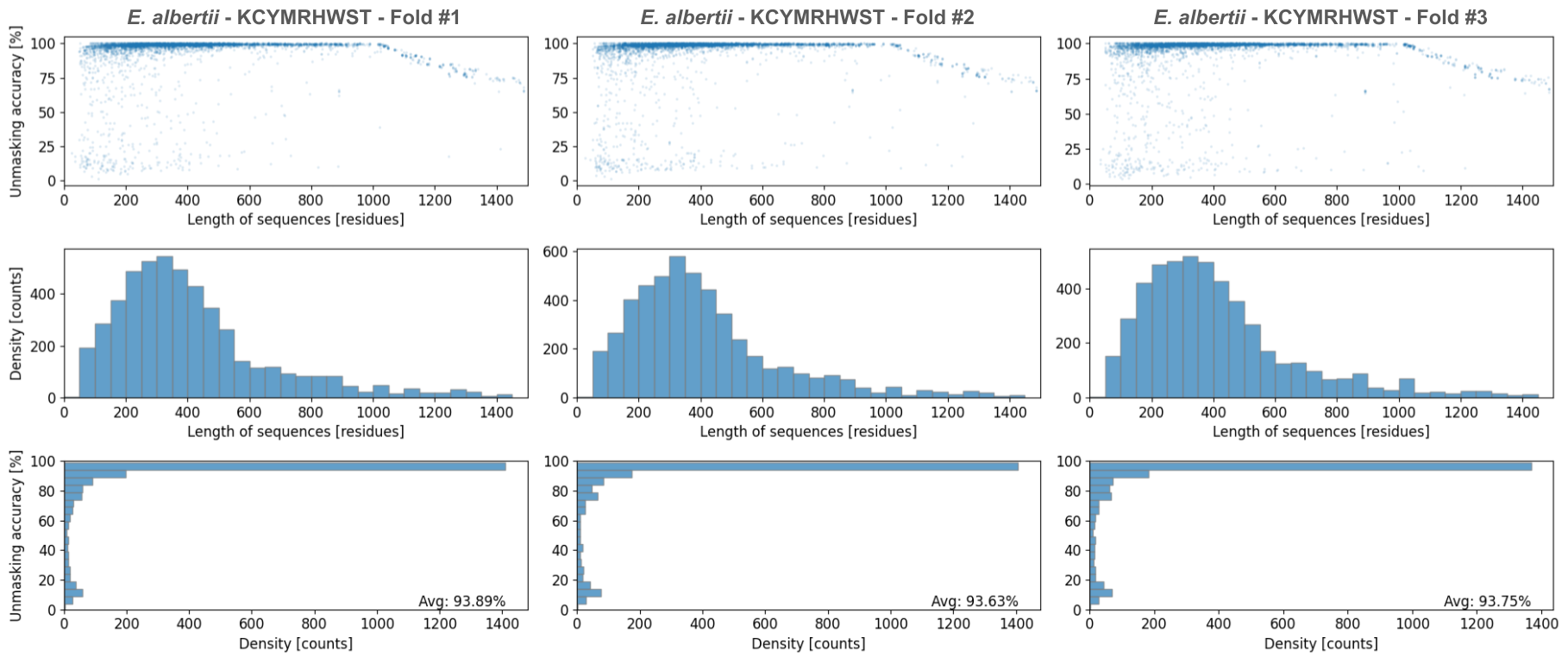}}
  \subfigure[\textit{E. fergusonii}]{\includegraphics[width=0.75\textwidth]{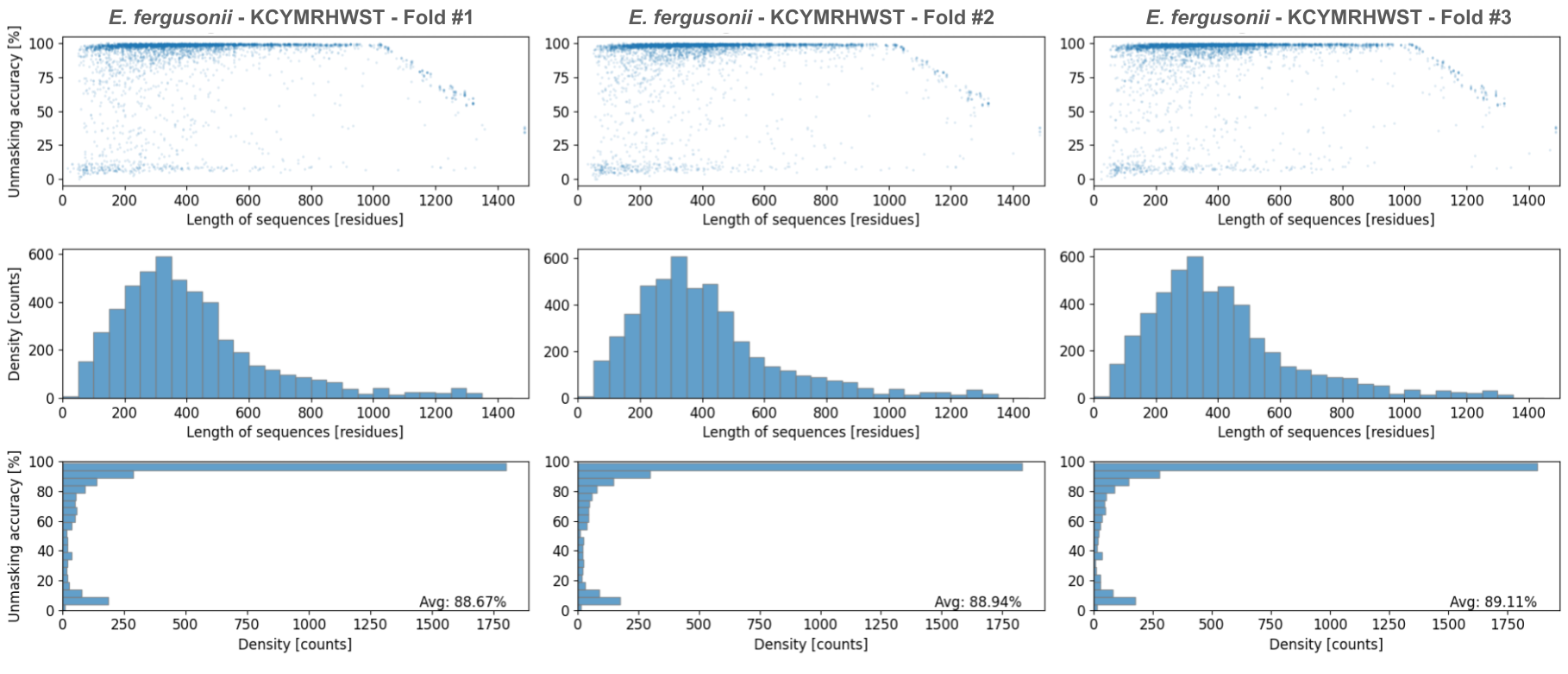}}
  \caption{The 3-fold unmasking inference results with density histograms from (a) \textit{E. coli}, (b) \textit{E. albertii}, and (c) \textit{E. fergusonii} with 9 known amino acids KCYMRHWST.}
  \label{sup:per-fold-KCYM}
  \Description{}
\end{figure*}

\newpage
\clearpage
\onecolumn
\vspace{15pt}

ID = A0A0H3PHF4

Description = Probable csgAB operon transcriptional regulatory protein n=32 RepID=A0A0H3PHF4\_ECO5C

Length = 240 (aa)

Predicted matches = 160 / 211 (75.83\%)

TM-score = 0.6319

Superposition in the TM-score: Length(d<5.0)= 143

\begin{figure}[h]
\centering
\includegraphics[width=\textwidth]{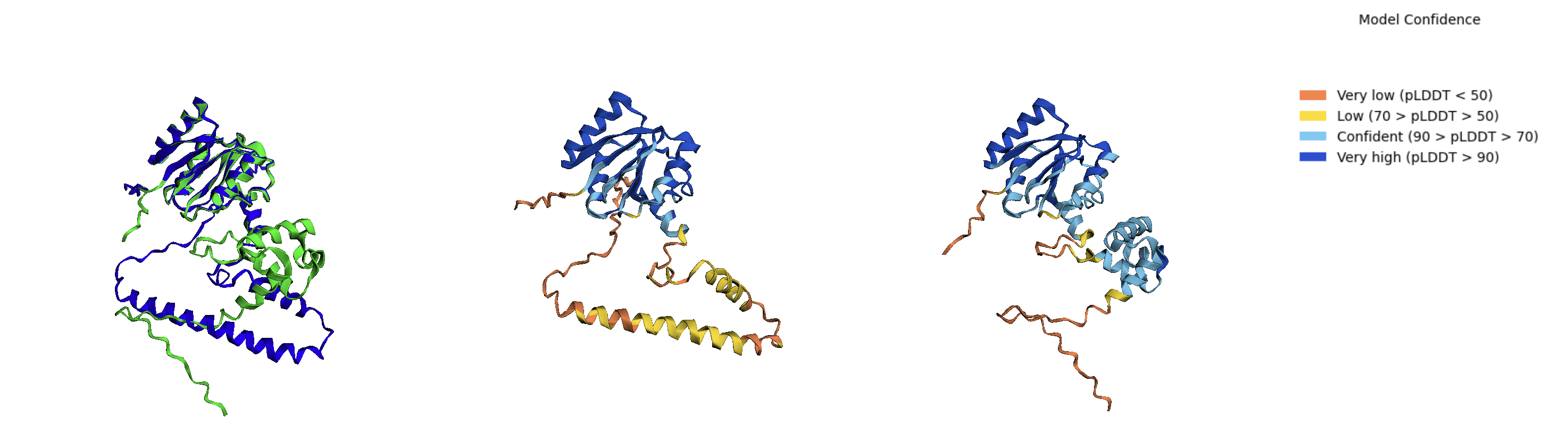}
\caption{Left: AlphaFold structures of predicted sequence (blue) and of UniProtKB (green). Center: AlphaFold structure of predicted sequence. Right: AlphaFold structure of UniProtKB sequence.}
\end{figure}

Alignment between predicted sequences and UniProtLB sequence:
 {
\begin{spacing}{0.5}
\ttfamily
\fontsize{7}{11}\selectfont
\texttt{}\\\textcolor{black}{M}\textcolor{green}{F}\textcolor{green}{N}\textcolor{green}{E}\textcolor{green}{V}\textcolor{green}{H}\textcolor{green}{S}\textcolor{green}{I}\textcolor{green}{H}\textcolor{green}{G\ }\textcolor{green}{H}\textcolor{green}{T}\textcolor{green}{L}\textcolor{green}{L}\textcolor{green}{L}\textcolor{green}{I}\textcolor{green}{T}\textcolor{black}{K}\textcolor{green}{P}\textcolor{green}{S\ }\textcolor{green}{L}\textcolor{green}{Q}\textcolor{green}{A}\textcolor{green}{T}\textcolor{green}{A}\textcolor{green}{L}\textcolor{green}{L}\textcolor{green}{Q}\textcolor{green}{H}\textcolor{green}{L\ }\textcolor{black}{K}\textcolor{red}{H}\textcolor{green}{S}\textcolor{green}{L}\textcolor{green}{A}\textcolor{green}{I}\textcolor{green}{T}\textcolor{green}{G}\textcolor{black}{K}\textcolor{green}{L\ }\textcolor{green}{H}\textcolor{green}{N}\textcolor{green}{I}\textcolor{green}{Q}\textcolor{green}{R}\textcolor{green}{S}\textcolor{green}{L}\textcolor{green}{D}\textcolor{green}{D}\textcolor{green}{I\ }\textcolor{green}{S}\textcolor{green}{S}\textcolor{green}{G}\textcolor{green}{S}\textcolor{green}{I}\textcolor{green}{I}\textcolor{red}{I}\textcolor{red}{V}\textcolor{green}{D}\textcolor{black}{M\ }\textcolor{black}{M}\textcolor{green}{E}\textcolor{green}{A}\textcolor{green}{D}\textcolor{black}{K}\textcolor{black}{K}\textcolor{green}{L}\textcolor{green}{I}\textcolor{green}{H}\textcolor{black}{Y\ }\textcolor{green}{W}\textcolor{green}{Q}\textcolor{green}{D}\textcolor{green}{T}\textcolor{green}{L}\textcolor{green}{S}\textcolor{green}{R}\textcolor{black}{K}\textcolor{green}{N}\textcolor{green}{N\ }\textcolor{green}{N}\textcolor{green}{I}\textcolor{black}{K}\textcolor{green}{I}\textcolor{green}{L}\textcolor{green}{L}\textcolor{green}{L}\textcolor{green}{N}\textcolor{green}{T}\textcolor{green}{P\ }\textcolor{green}{E}\textcolor{green}{D}\textcolor{black}{Y}\textcolor{green}{P}\textcolor{black}{Y}\textcolor{green}{R}\textcolor{green}{D}\textcolor{green}{I}\textcolor{green}{E}\textcolor{green}{N\ }\textcolor{green}{W}\textcolor{green}{P}\textcolor{green}{H}\textcolor{green}{I}\textcolor{green}{N}\textcolor{green}{G}\textcolor{green}{V}\textcolor{green}{F}\textcolor{black}{Y}\textcolor{green}{A\ }\\
\textcolor{black}{\ }\textcolor{black}{\ }\textcolor{black}{\ }\textcolor{black}{\ }\textcolor{black}{\ }\textcolor{black}{\ }\textcolor{black}{:}\textcolor{black}{:}\textcolor{black}{:}\textcolor{black}{:\ }\textcolor{black}{:}\textcolor{black}{:}\textcolor{black}{:}\textcolor{black}{:}\textcolor{black}{:}\textcolor{black}{:}\textcolor{black}{:}\textcolor{black}{:}\textcolor{black}{:}\textcolor{black}{:\ }\textcolor{black}{:}\textcolor{black}{:}\textcolor{black}{:}\textcolor{black}{:}\textcolor{black}{:}\textcolor{black}{:}\textcolor{black}{:}\textcolor{black}{:}\textcolor{black}{:}\textcolor{black}{:\ }\textcolor{black}{:}\textcolor{black}{:}\textcolor{black}{:}\textcolor{black}{:}\textcolor{black}{:}\textcolor{black}{:}\textcolor{black}{:}\textcolor{black}{:}\textcolor{black}{:}\textcolor{black}{:\ }\textcolor{black}{:}\textcolor{black}{:}\textcolor{black}{:}\textcolor{black}{:}\textcolor{black}{:}\textcolor{black}{:}\textcolor{black}{:}\textcolor{black}{:}\textcolor{black}{:}\textcolor{black}{:\ }\textcolor{black}{:}\textcolor{black}{:}\textcolor{black}{:}\textcolor{black}{:}\textcolor{black}{:}\textcolor{black}{:}\textcolor{black}{:}\textcolor{black}{:}\textcolor{black}{:}\textcolor{black}{:\ }\textcolor{black}{:}\textcolor{black}{:}\textcolor{black}{:}\textcolor{black}{:}\textcolor{black}{:}\textcolor{black}{:}\textcolor{black}{:}\textcolor{black}{:}\textcolor{black}{:}\textcolor{black}{:\ }\textcolor{black}{:}\textcolor{black}{:}\textcolor{black}{:}\textcolor{black}{:}\textcolor{black}{:}\textcolor{black}{:}\textcolor{black}{:}\textcolor{black}{:}\textcolor{black}{:}\textcolor{black}{:\ }\textcolor{black}{:}\textcolor{black}{:}\textcolor{black}{:}\textcolor{black}{:}\textcolor{black}{:}\textcolor{black}{:}\textcolor{black}{:}\textcolor{black}{:}\textcolor{black}{:}\textcolor{black}{:\ }\textcolor{black}{:}\textcolor{black}{:}\textcolor{black}{:}\textcolor{black}{:}\textcolor{black}{:}\textcolor{black}{:}\textcolor{black}{:}\textcolor{black}{:}\textcolor{black}{:}\textcolor{black}{:\ }\textcolor{black}{:}\textcolor{black}{:}\textcolor{black}{:}\textcolor{black}{:}\textcolor{black}{:}\textcolor{black}{:}\textcolor{black}{:}\textcolor{black}{:}\textcolor{black}{:}\textcolor{black}{:\ }\\
\textcolor{black}{M}\textcolor{green}{F}\textcolor{green}{N}\textcolor{green}{E}\textcolor{green}{V}\textcolor{green}{H}\textcolor{green}{S}\textcolor{green}{I}\textcolor{green}{H}\textcolor{green}{G\ }\textcolor{green}{H}\textcolor{green}{T}\textcolor{green}{L}\textcolor{green}{L}\textcolor{green}{L}\textcolor{green}{I}\textcolor{green}{T}\textcolor{black}{K}\textcolor{green}{P}\textcolor{green}{S\ }\textcolor{green}{L}\textcolor{green}{Q}\textcolor{green}{A}\textcolor{green}{T}\textcolor{green}{A}\textcolor{green}{L}\textcolor{green}{L}\textcolor{green}{Q}\textcolor{green}{H}\textcolor{green}{L\ }\textcolor{black}{K}\textcolor{red}{Q}\textcolor{green}{S}\textcolor{green}{L}\textcolor{green}{A}\textcolor{green}{I}\textcolor{green}{T}\textcolor{green}{G}\textcolor{black}{K}\textcolor{green}{L\ }\textcolor{green}{H}\textcolor{green}{N}\textcolor{green}{I}\textcolor{green}{Q}\textcolor{green}{R}\textcolor{green}{S}\textcolor{green}{L}\textcolor{green}{D}\textcolor{green}{D}\textcolor{green}{I\ }\textcolor{green}{S}\textcolor{green}{S}\textcolor{green}{G}\textcolor{green}{S}\textcolor{green}{I}\textcolor{green}{I}\textcolor{red}{L}\textcolor{red}{L}\textcolor{green}{D}\textcolor{black}{M\ }\textcolor{black}{M}\textcolor{green}{E}\textcolor{green}{A}\textcolor{green}{D}\textcolor{black}{K}\textcolor{black}{K}\textcolor{green}{L}\textcolor{green}{I}\textcolor{green}{H}\textcolor{black}{Y\ }\textcolor{green}{W}\textcolor{green}{Q}\textcolor{green}{D}\textcolor{green}{T}\textcolor{green}{L}\textcolor{green}{S}\textcolor{green}{R}\textcolor{black}{K}\textcolor{green}{N}\textcolor{green}{N\ }\textcolor{green}{N}\textcolor{green}{I}\textcolor{black}{K}\textcolor{green}{I}\textcolor{green}{L}\textcolor{green}{L}\textcolor{green}{L}\textcolor{green}{N}\textcolor{green}{T}\textcolor{green}{P\ }\textcolor{green}{E}\textcolor{green}{D}\textcolor{black}{Y}\textcolor{green}{P}\textcolor{black}{Y}\textcolor{green}{R}\textcolor{green}{D}\textcolor{green}{I}\textcolor{green}{E}\textcolor{green}{N\ }\textcolor{green}{W}\textcolor{green}{P}\textcolor{green}{H}\textcolor{green}{I}\textcolor{green}{N}\textcolor{green}{G}\textcolor{green}{V}\textcolor{green}{F}\textcolor{black}{Y}\textcolor{green}{A\ }\\
\\
\textcolor{black}{M}\textcolor{green}{E}\textcolor{green}{D}\textcolor{green}{Q}\textcolor{green}{E}\textcolor{green}{R}\textcolor{green}{V}\textcolor{green}{V}\textcolor{green}{N}\textcolor{green}{G\ }\textcolor{green}{L}\textcolor{green}{Q}\textcolor{green}{G}\textcolor{green}{V}\textcolor{green}{L}\textcolor{green}{R}\textcolor{green}{G}\textcolor{green}{E}\textcolor{black}{C}\textcolor{black}{Y\ }\textcolor{green}{F}\textcolor{green}{T}\textcolor{green}{Q}\textcolor{black}{K}\textcolor{green}{L}\textcolor{green}{A}\textcolor{green}{S}\textcolor{black}{Y}\textcolor{green}{L}\textcolor{green}{I\ }\textcolor{green}{T}\textcolor{green}{H}\textcolor{green}{S}\textcolor{green}{G}\textcolor{green}{N}\textcolor{black}{Y}\textcolor{green}{R}\textcolor{black}{Y}\textcolor{green}{N}\textcolor{green}{S\ }\textcolor{green}{T}\textcolor{green}{E}\textcolor{green}{S}\textcolor{green}{A}\textcolor{green}{L}\textcolor{green}{L}\textcolor{red}{N}\textcolor{green}{H}\textcolor{green}{R}\textcolor{green}{E\ }\textcolor{black}{K}\textcolor{red}{P}\textcolor{green}{I}\textcolor{green}{L}\textcolor{red}{E}\textcolor{black}{K}\textcolor{green}{L}\textcolor{green}{R}\textcolor{green}{I}\textcolor{red}{L\ }\textcolor{green}{A}\textcolor{green}{S}\textcolor{green}{N}\textcolor{green}{N}\textcolor{red}{V}\textcolor{green}{I}\textcolor{green}{A}\textcolor{red}{D}\textcolor{red}{T}\textcolor{red}{S\ }\textcolor{green}{F}\textcolor{red}{F}\textcolor{red}{I}\textcolor{green}{E}\textcolor{red}{Q}\textcolor{red}{I}\textcolor{green}{V}\textcolor{black}{K}\textcolor{red}{G}\textcolor{green}{H\ }\textcolor{green}{L}\textcolor{black}{Y}\textcolor{red}{V}\textcolor{green}{L}\textcolor{green}{F}\textcolor{black}{K}\textcolor{black}{K}\textcolor{green}{I}\textcolor{red}{V}\textcolor{red}{N\ }\textcolor{black}{K}\textcolor{red}{S}\textcolor{green}{R}\textcolor{red}{E}\textcolor{red}{R}\textcolor{green}{A}\textcolor{red}{A}\textcolor{red}{I}\textcolor{red}{L}\textcolor{red}{G\ }\textcolor{red}{L}\textcolor{red}{T}\textcolor{red}{R}\textcolor{green}{S}\textcolor{green}{A}\textcolor{red}{D}\textcolor{green}{S}\textcolor{red}{T}\textcolor{red}{L}\textcolor{red}{I\ }\\
\textcolor{black}{:}\textcolor{black}{:}\textcolor{black}{:}\textcolor{black}{:}\textcolor{black}{:}\textcolor{black}{:}\textcolor{black}{:}\textcolor{black}{:}\textcolor{black}{:}\textcolor{black}{:\ }\textcolor{black}{:}\textcolor{black}{:}\textcolor{black}{:}\textcolor{black}{:}\textcolor{black}{:}\textcolor{black}{:}\textcolor{black}{:}\textcolor{black}{:}\textcolor{black}{:}\textcolor{black}{:\ }\textcolor{black}{:}\textcolor{black}{:}\textcolor{black}{:}\textcolor{black}{:}\textcolor{black}{:}\textcolor{black}{:}\textcolor{black}{:}\textcolor{black}{:}\textcolor{black}{:}\textcolor{black}{:\ }\textcolor{black}{:}\textcolor{black}{:}\textcolor{black}{:}\textcolor{black}{:}\textcolor{black}{:}\textcolor{black}{:}\textcolor{black}{\ }\textcolor{black}{\ }\textcolor{black}{\ }\textcolor{black}{\ \ }\textcolor{black}{\ }\textcolor{black}{\ }\textcolor{black}{\ }\textcolor{black}{:}\textcolor{black}{\ }\textcolor{black}{\ }\textcolor{black}{\ }\textcolor{black}{\ }\textcolor{black}{\ }\textcolor{black}{\ \ }\textcolor{black}{\ }\textcolor{black}{\ }\textcolor{black}{\ }\textcolor{black}{\ }\textcolor{black}{:}\textcolor{black}{\ }\textcolor{black}{\ }\textcolor{black}{\ }\textcolor{black}{:}\textcolor{black}{\ \ }\textcolor{black}{\ }\textcolor{black}{\ }\textcolor{black}{\ }\textcolor{black}{\ }\textcolor{black}{\ }\textcolor{black}{\ }\textcolor{black}{\ }\textcolor{black}{\ }\textcolor{black}{\ }\textcolor{black}{\ \ }\textcolor{black}{\ }\textcolor{black}{\ }\textcolor{black}{\ }\textcolor{black}{\ }\textcolor{black}{\ }\textcolor{black}{\ }\textcolor{black}{\ }\textcolor{black}{\ }\textcolor{black}{\ }\textcolor{black}{\ \ }\textcolor{black}{\ }\textcolor{black}{\ }\textcolor{black}{\ }\textcolor{black}{\ }\textcolor{black}{\ }\textcolor{black}{\ }\textcolor{black}{\ }\textcolor{black}{\ }\textcolor{black}{\ }\textcolor{black}{\ \ }\textcolor{black}{\ }\textcolor{black}{\ }\textcolor{black}{\ }\textcolor{black}{\ }\textcolor{black}{\ }\textcolor{black}{\ }\textcolor{black}{\ }\textcolor{black}{\ }\textcolor{black}{\ }\textcolor{black}{\ \ }\textcolor{black}{\ }\textcolor{black}{\ }\textcolor{black}{\ }\textcolor{black}{\ }\textcolor{black}{\ }\textcolor{black}{\ }\textcolor{black}{\ }\textcolor{black}{\ }\textcolor{black}{\ }\textcolor{black}{\ \ }\\
\textcolor{black}{M}\textcolor{green}{E}\textcolor{green}{D}\textcolor{green}{Q}\textcolor{green}{E}\textcolor{green}{R}\textcolor{green}{V}\textcolor{green}{V}\textcolor{green}{N}\textcolor{green}{G\ }\textcolor{green}{L}\textcolor{green}{Q}\textcolor{green}{G}\textcolor{green}{V}\textcolor{green}{L}\textcolor{green}{R}\textcolor{green}{G}\textcolor{green}{E}\textcolor{black}{C}\textcolor{black}{Y\ }\textcolor{green}{F}\textcolor{green}{T}\textcolor{green}{Q}\textcolor{black}{K}\textcolor{green}{L}\textcolor{green}{A}\textcolor{green}{S}\textcolor{black}{Y}\textcolor{green}{L}\textcolor{green}{I\ }\textcolor{green}{T}\textcolor{green}{H}\textcolor{green}{S}\textcolor{green}{G}\textcolor{green}{N}\textcolor{black}{Y}\textcolor{green}{R}\textcolor{black}{Y}\textcolor{green}{N}\textcolor{green}{S\ }\textcolor{green}{T}\textcolor{green}{E}\textcolor{green}{S}\textcolor{green}{A}\textcolor{green}{L}\textcolor{green}{L}\textcolor{red}{T}\textcolor{green}{H}\textcolor{green}{R}\textcolor{green}{E\ }\textcolor{black}{K}\textcolor{red}{E}\textcolor{green}{I}\textcolor{green}{L}\textcolor{red}{N}\textcolor{black}{K}\textcolor{green}{L}\textcolor{green}{R}\textcolor{green}{I}\textcolor{red}{G\ }\textcolor{green}{A}\textcolor{green}{S}\textcolor{green}{N}\textcolor{green}{N}\textcolor{red}{E}\textcolor{green}{I}\textcolor{green}{A}\textcolor{red}{R}\textcolor{red}{S}\textcolor{red}{L\ }\textcolor{green}{F}\textcolor{red}{I}\textcolor{red}{S}\textcolor{green}{E}\textcolor{red}{N}\textcolor{red}{T}\textcolor{green}{V}\textcolor{black}{K}\textcolor{red}{T}\textcolor{green}{H\ }\textcolor{green}{L}\textcolor{black}{Y}\textcolor{red}{N}\textcolor{green}{L}\textcolor{green}{F}\textcolor{black}{K}\textcolor{black}{K}\textcolor{green}{I}\textcolor{red}{A}\textcolor{red}{V\ }\textcolor{black}{K}\textcolor{red}{N}\textcolor{green}{R}\textcolor{red}{T}\textcolor{red}{Q}\textcolor{green}{A}\textcolor{red}{V}\textcolor{red}{S}\textcolor{red}{W}\textcolor{red}{A\ }\textcolor{red}{N}\textcolor{red}{D}\textcolor{red}{T}\textcolor{green}{S}\textcolor{green}{A}\textcolor{red}{I}\textcolor{green}{S}\textcolor{red}{H}\textcolor{red}{E}\textcolor{red}{T\ }\\
\\
\textcolor{red}{A}\textcolor{red}{G}\textcolor{red}{N}\textcolor{red}{N}\textcolor{red}{P}\textcolor{black}{C}\textcolor{red}{S}\textcolor{red}{V}\textcolor{red}{S}\textcolor{red}{A\ }\textcolor{green}{S}\textcolor{red}{S}\textcolor{red}{S}\textcolor{red}{S}\textcolor{red}{A}\textcolor{red}{G}\textcolor{red}{A}\textcolor{red}{T}\textcolor{red}{L}\textcolor{red}{W\ }\\
\textcolor{black}{\ }\textcolor{black}{\ }\textcolor{black}{\ }\textcolor{black}{\ }\textcolor{black}{\ }\textcolor{black}{\ }\textcolor{black}{\ }\textcolor{black}{\ }\textcolor{black}{\ }\textcolor{black}{\ \ }\textcolor{black}{\ }\textcolor{black}{\ }\textcolor{black}{\ }\textcolor{black}{\ }\textcolor{black}{\ }\textcolor{black}{\ }\textcolor{black}{\ }\textcolor{black}{\ }\textcolor{black}{\ }\textcolor{black}{\ \ }\\
\textcolor{red}{L}\textcolor{red}{F}\textcolor{red}{T}\textcolor{red}{L}\textcolor{red}{D}\textcolor{black}{C}\textcolor{red}{G}\textcolor{red}{G}\textcolor{red}{R}\textcolor{red}{I\ }\textcolor{green}{S}\textcolor{red}{V}\textcolor{red}{R}\textcolor{red}{R}\textcolor{red}{R}\textcolor{red}{E}\textcolor{red}{S}\textcolor{red}{S}\textcolor{red}{R}\textcolor{red}{R\ }\\
\\
\end{spacing}
}

\hrule

\end{document}